\documentclass[draft]{aa}

\newcommand{\etal}{{\it et al.}}
\newcommand{\kms}{\mbox{\ km\ s$^{-1}$}}
\newcommand{\msunyr}{\mbox{M$_{\odot}$\thinspace yr$^{-1}\;$}}
\newcommand{\msun}{\mbox{$M_{\odot}\;$}}
\newcommand{\mstar}{\mbox{$M_{\ast}\;$}}
\newcommand{\rsun}{\mbox{$R_{\odot}\;$}}
\newcommand{\rstar}{\mbox{$R_{\ast}\;$}}

\newcommand{\mnras}{{\sl MNRAS}}
\newcommand{\apj}{{\sl ApJ}}

\newcommand{\apjs}{{\sl ApJS}}

\newcommand{\aanda}{{\sl A\&A}}
\newcommand{\mdot}{\mbox{$\stackrel{.}{\textstyle M}$}}

\title{On the decoupling and reaccretion of low density, line-driven winds}
\titlerunning{Reaccretion of line-driven winds}
\author{John M. Porter \and Barry A. Skouza}
\institute{Astrophysics Research Institute, 
Liverpool John Moores University, Twelve Quays House, Egerton Wharf, \\
Birkenhead. L41 1LD, UK
\thanks{email : {\tt jmp@astro.livjm.ac.uk} }
}

\date{}

\begin{document}
\maketitle

\begin{abstract}
The flow generated by low-density radiatively driven winds which
decouple their gas and radiation fields is discussed. In particular we
concentrate on flow which is still bound to the star and can therefore
reaccrete. The wind decelerates after decoupling and eventually stalls.
A shell of gas
is generated, and we find that this shell is unstable
and contracts back to the star with periods of hours to days.

We find that the pulsating shells may be difficult to observe, as
their emission is variable and the maximum emission at H$_\alpha$ 
(of $\sim 1\%$ of the continuum) occurs over a small fraction of the
shell cycle. 


\end{abstract}

%-----------------------------------------------------------------------%
\section{Introduction}

The winds from hot stars are thought to be generated by radiation
pressure on optically thick UV resonance lines, 
and the theory of line-driven flow is very successful in 
accounting many observed wind features (see Castor, Abbott \& Klein 1975,
Abbott 1980, Pauldrach \etal\ 1986, Kudritzki \etal\ 1989). 

In low density winds, however, there exists the possibility that 
the radiation force and the wind flow may decouple (Springmann \&
Pauldrach 1992, Porter \& Drew 1995 [hereafter PD95], Babel 1996).
The decoupling process strips the metallic ions from the rest of the
plasma, and as the radiative force on the flow is mediated by the
ions, then the wind receives no further acceleration.
The winds which are most likely to undergo this decoupling are B star
winds (Babel 1996, PD95), and metallic A star winds (Babel 1995).
The frictional interaction between the metallic ions and the rest of
the wind may also seriously interfere with the radiative equilibrium
(Springmann \& Pauldrach 1992, Gayley \& Owocki 1995).
Decoupling radii for low density winds may be close to
the photosphere for B stars (PD95) or indeed may be associated with the
photosphere on the case of some A star winds (Babel 1995), where
there is no region outside the photosphere where a fully coupled wind
exists.

The line-driven wind accelerates normally when fully coupled to the
radiation field, but once decoupled cannot receive any further
acceleration. 
It is possible that the wind will decouple before it has reached
escape velocity, in which case the decoupled flow will stall at some
radius and fall back toward the star. It is this aspect of decoupled
flows which is examined in this paper.

The rest of this paper is structured as follows: in \S2 the
physics of wind decoupling is examined, and 
in \S3 hydrodynamical simulations of decoupled winds are presented.
The observational signatures of the shells are presented in \S4 and 
a discussion and conclusions are given in \S5.  

%-----------------------------------------------------------------------%
\section{Decoupling physics}

There are two ways in which the radiation and matter fields in a
line-driven wind may decouple: ion stripping (PD95, Springmann \&
Pauldrach 1992) and shock decoupling (PD95, Krolik \& Raymond 1985). 

The physics of ion stripping 
was first noted in the context of electrical conductivity 
by Dreicer (1959, 1960) and has its roots in the basic mechanism
allowing optically thick metallic ion lines to mediate the force on a
wind. 
The metallic ions are accelerated via photon scattering off their UV
lines. The ions then share this acceleration with the rest of the wind
(hydrogen and helium ions) by a process similar to friction. This
frictional interaction depends on the relative drift velocity of the
ions through the rest of the wind. For low drift velocities the force
is proportional to the drift velocity, however it reaches a maximum
when the drift and thermal kinetic energies are equal. Beyond this the
frictional interaction decreases rapidly with increasing drift
velocity. Therefore if the ions' drift
velocity does become large enough, then as there is little or no
frictional interaction with the rest of the wind, they may freely
accelerate. This leaves the rest of the wind with no acceleration, and
it will then just be acted upon by the gravitational attraction of the
central star.

Shock decoupling occurs in a different way. It has been known that hot
star winds may be unstable to the growth of instabilities for some
time (Lucy \& Solomon 1970, MacGregor \etal\ 1979, Owocki \& Rybicki
1984) although it is unclear whether the winds are inherently unstable
or only advectively unstable (see the review by Owocki 1991).
If a low density wind passes through a strong shock, then the
postshock gas may be too diffuse to cool radiatively, and so the gas
remains in its high temperature, superionized state (Krolik
\& Raymond 1985). The wind may decouple from the radiation if the
postshock gas remains out of equilibrium, as the main
metallic species responsible for line-driving have been ionized away
(Castor 1987).

Although both of these processes may in principle occur, only the
former (ion stripping) is considered here as it is more deterministic -- the
decoupling radius for shock decoupling can only be calculated if 
the star's X-ray emission (to determine the shock velocity)  and
mass-loss rate are available, whereas for 
ion stripping, only the mass-loss rate is required.
It should be noted that the decoupling radius is similar in each case
(see PD95). 

For ion stripping, the decoupling radius $r_d$ may be estimated by
considering the relative velocity between the ions and the rest of
the wind. As it is this drift velocity which sets the frictional
interaction then it also defines how much acceleration is imparted to
the rest of the wind. When the drift velocity becomes comparable to
the ions thermal speed, the frictional interaction reaches a
maximum beyond which it falls with increasing drift velocity. 
At this point the frictional
interaction between the ions and the rest of the wind becomes small
and the ions are (nearly) free to accelerate out of the wind. This
radius is where the matter and radiation fields effectively decouple.
For a beta-velocity law this yields
\begin{equation}
\frac{\beta\; v_{\infty\;3}^3 \;(1 - \rstar/r_d)^{3\beta - 1}
\;T_4 \;\rstar} 
{\dot{M}_{-9}} = 1.5\times 10^3 \;Y_i \;Z_i^2 \;\ln\Lambda,
\label{criteria}
\end{equation}
(see Springmann \& Pauldrach 1992 and also Gayley \& Owocki 1995)
where $v_{\infty\;3}$ is the terminal velocity of the wind measured in
units of $10^3$\kms, and the stellar
radius $\rstar$ is expressed in solar units. 
Also $Y_i$ is the mass fraction of the
ions, $Z_i$ is the degree of ionization, and $\dot{M}_{-9} \equiv \dot{M} /
10^{-9}$\msunyr is the mass loss rate derived from UV line profiles. $T_4$ is
the temperature of the gas in units of 10$^4$K and ln$\Lambda$ is the
Coulomb logarithm.
The approach of calculating the ion-stripping radius used to derive
eq.1 is, however, strictly only valid when the wind is fully coupled.

Therefore, a more exact calculation needs to be
undertaken including the precise frictional force expression 
(e.g. see Springmann \& Pauldrach 1992, Dreicer 1959, 1960).
Although the frictional heating may increase the temperature of the
wind and cause decoupling at smaller radii (PD95 noted this fact), 
the decoupling radii and velocities for a model B2 star have been
calculated assuming isothermality. 
Temperature profiles of coupled radiatively-driven winds have been
calculated by Drew (1989) using a full description of radiative heating and
cooling processes in the wind. Drew's calculations show that the wind
temperature falls below 0.6$T_{\rm eff}$ only for $r > 2\rstar\!$,
although the frictional heating from the relative drift of ions through
the rest of the wind was not included.
Therefore the assumptions made here that the wind
temperature is 0.8$T_{\rm eff}$ (Klein \& Castor 1978), and that the
flow is isothermal is probably not seriously in error.

For all the calculations undertaken in this paper, we use a
``standard'' B2V star 
which has stellar parameters of \mstar = 7.5\msun,
\rstar = 4.0\rsun, $T_{\rm eff} = 20,000$K.
The escape velocity of this star is $v_{\rm esc} = 840\kms$, and the 
fully coupled terminal wind velocity is $v_\infty =
2.2\alpha v_{\rm esc} /(1-\alpha) = 2260\kms$ (Friend \& Abbott
1986), for the line-driving parameter $\alpha = 0.55$ (see Castor,
Abbott \& Klein 1975).
The decoupling radii and velocities have been calculated for our
standard star, and are presented as a function of mass-loss rate in
fig.1. 

\begin{figure}
\begin{picture}(200,140)
\put(0,0){\includegraphics{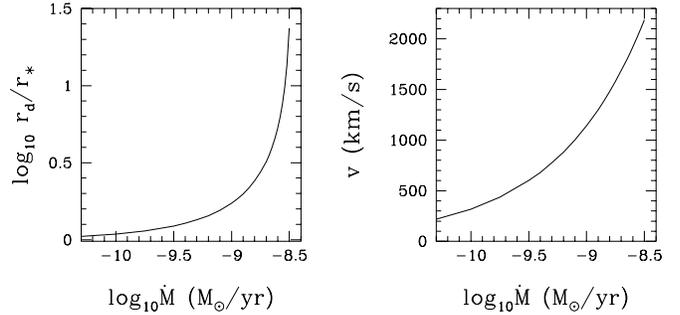}}
\end{picture}
\caption{Decoupling radii (left) and velocity (right) for
a 7.5\msun, 4.0\rsun, 2$\times 10^4$K star as a function of mass-loss
rate. The escape velocity is 840\kms\ and the wind terminal velocity
is 2260\kms.}
\end{figure}
The escape velocity of the star is 840\kms\ and so 
from fig.1, it can be seen that if the mass-loss rate is less than
$\sim 10^{-9.2}$\msunyr ($=6.3\times 10^{-10}$\msunyr), then the wind
will decouple and {\em still} be bound to the star.

%-----------------------------------------------------------------------
\section{Results}

\subsection{Analytic considerations}
If the radiation and matter fields decouple at a velocity
$v_d < v_{\rm esc}$ then, as noted above, the material is still bound
to the star. 
In this case the local radial velocity must be calculated for flow in
a gravitational field. The radial velocity will decrease until the
flow stalls at an outer radius $r_s$ which can be estimated by
equating the change in potential and kinetic energies, leading to
\begin{equation}
\frac{r_s}{\rstar}
\approx \left[ \frac{\rstar}{r_d} - \frac{1}{2}
\left(\frac{\rstar a^2}{G\mstar}\right) \left( \frac{v_d}{a} \right)^2
\right]^{-1}.
\end{equation}
As there is flow
still with positive radial momentum behind this stalling front, a
shell of stalled wind will form being supported by the ram pressure of
the gas behind it.
If we assume that the shell {\em can} 
be supported by the wind at some radius $r_{\rm shell}$, then we can
estimate the 
maximum total mass in the shell. 
The supporting ram-pressure $P_{\rm ram}$
of the stalling gas will be $\rho v^2$ 
where $\rho$ is the density and $v$ is the local radial velocity.
The oppositely directed pressure generated by the shell is the due to
the attractive 
gravitational force of the star $G\mstar M_{\rm shell} /r_{\rm shell}^2$ where
$M_{\rm shell}$ is the shell mass and $G$ is the gravitational
constant. This force acts over the surface area of the shell $4\pi
r_{\rm shell}^2$.
Balancing the two oppositely directed pressures
yields
\begin{equation}
M_{\rm shell} \approx \frac{\mdot v r_{\rm shell}^2}{G\mstar}.
\end{equation}
Here we have used the mass-continuity equation for spherical flows
$\mdot = 4\pi r^2 \rho v$, where \mdot\ is the mass-loss rate of the wind.
If $v_2 = v / 100\kms$ and \mdot$_{-9} = \mdot / 10^{-9}\msunyr$, then
\begin{equation}
\frac{M_{\rm shell}}\msun \approx 1.2\times 10^{-14} \mdot_{-9} v_2
\left( \frac{r_s}{\rsun} \right)^2 \left( \frac{\mstar}{\msun} \right)^{-1}.
\end{equation}
Note that 
a distinction has been made between the stalling radius $r_s$ and the
radius of the shell $r_{\rm shell}$. This is because the shell will
only appear at the 
stalling radius if the wind first expands out into a vacuum. If there
is any ambient density around the star $\rho_{\rm amb}$, then this is
swept up into the shell when the wind is ``turned on''.

The evolution of the wind will broadly be composed of
several parts.
Initially the wind will be driven into the surrounding medium and will
sweep up a shell of ambient gas. This shell will stop expanding when
the shell and ram pressures balance. The shell then grows as it is
receiving gas from the wind (at presumably the same rate as the
mass-loss rate from the star). Finally the shell becomes too massive
to be supported and will fall back toward the star.

If we assume that $\rho_{\rm amb}$ is small then the shell radius {\em is}
the stalling radius. Also the wind's ram pressure in this case is
the thermal pressure $\rho a^2$, where $a$ is
the sound speed.
For stalling radii of $r_s \approx 5\rstar = 20\rsun$
the shell mass for our standard star with a mass-loss rate of
$10^{-9.5}$\msunyr which will be supported by the stalling wind is 
$M_{\rm shell} \sim 3\times 10^{-14}\msun$. 
Given that the shell mass grows at a rate of \mdot, 
we find that it is stable for a time
\begin{equation}
t = 380 v_2 \left(\frac{r_s}{\rsun}\right)^2 \left(
\frac{\mstar}{\msun}\right)^{-1} \ {\rm s}
\end{equation}
which for the example above is $t \approx 1$hr.
The timescales here are similar to the flow timescale of
the wind (approximately $\rstar/v \approx 1$hr for a wind speed of
1000\kms). 

After this time the shell's mass will have increased such that  
there is insufficient ram pressure to levitate the shell and so it
will fall back to the star.
As the shell falls inward, the wind ram 
pressure increases - both the wind density and the radial
velocity $v$ increase (Mach numbers in radiatively driven flows can be
$\sim 100$). However, the increasing mass of the shell is now acting over a
smaller area. Hence the effective pressure exerted by the shell inward
due to the gravitational attraction by the central star increases.
This pressure of the shell is the gravitational force divided by the
area of the shell: $P_{\rm shell} \propto r^{-4}$ whereas the ram
pressure is $P_{\rm ram}\propto v/r^2$.
The ratio of these two pressures is
\begin{equation}
\frac{P_{\rm shell}}{P_{\rm ram}} = 
\frac{8.6\times 10^{-12}}{\mdot_{-9} v_2} 
\left( \frac{\mstar}{\msun} \right)
\left( \frac{M_{\rm shell}}{\msun} \right)
\left( \frac{r}{\rsun} \right)^{-2}.
\end{equation}
Note that in the decoupled part of the wind $v \sim r^{-1/2}$, and so 
as $M_{\rm shell}$ increases and moves toward the star ($r$ decreases)
the ratio above grows. This then implies that the shell can not be
stable (balanced) at any radius, and must fall back to the star.

Once the shell has contracted to the star, the wind is free again to
expand to its stalling point and the cycle restarts. 
This analysis predicts that stars with bound decoupled winds will
inevitably produce periodic structures associated with the wind.
However, a word of caution
is warranted here : this scenario predicts that a shell of gas should
form and collapse onto the star with timescales of
hours--days. There is a possibility that the oppositely directed
gravity and pressure gradients as the shell contracts may break up the shell
due to Rayleigh-Taylor instabilities.

\subsection{Numerical hydrodynamic modelling}
To illustrate the preceeding section's scenario, a 1D numerical
hydrodynamic simulation has been completed. The computer code
utilises the second order Godonuv scheme due to Falle (1991) to solve
the hydrodynamic equations. The line force is calculated using the standard
Castor, Abbott \& Klein (1976) formalism of the force multiplier,
supplemented with the finite-disc correction (Friend \& Abbott 1986)
and the density weighting factor from Abbott (1982).
The grid has a spacing of 0.005\rstar at $r =$\rstar, increasing by
1.5\% to a 
maximum spacing of 0.09\rstar at the outer radius of 6.75\rstar
(there are 200 grid points).
The outer boundary allows gas to flow through it freely, whereas the
inner boundary has the density fixed at $10^{-12}$g cm$^{-3}$, and the
velocity extrapolated from the neighbouring zones. 
We can expect that after a time the radial velocity at the outer
boundary will become negative and matter will flow into the
computational domain. This causes the density to rise at the outer
parts of the simulation, and finally lead to a significant ram
pressure acting on the shell from infalling material. Ultimately this
leads to the central star being surrounded by high density, infalling
material, completely thwarting even the initiation of the wind!
Clearly this is an artifact of the outer boundary.
After experimentation, it is found that the outer boundary which illustrates
the scenario most effectively is a ``valve'', i.e. 
the boundary is free-flow when the radial velocity is greater than
zero, but otherwise we set the radial velocity to zero, and the density to a
small value ($10^{-30}$g cm$^{-3}$).

In order to examine the effect of differing mass-loss rates on the
flow we prepared five fully-coupled wind solutions as initial
conditions. These have been generated for our standard B2 star. We
find that the mass-loss rate for this star
is $\sim 10^{-8.3}$\msunyr. As we require differing (and much lower)
mass-loss rates to assess its effect on the decoupled flows, 
we have reduced the radiation force by as much as an order of
magnitude. This may be seen as a change in metallicity -- the
radiative acceleration is proportional to the metallicity raised to
the power of 
$1 -\alpha$ (Abbott 1982, Castor, Abbott \& Klein 1975). The lowest
metallicity model implied is then 0.006 solar. 
By reducing the line force we have generated
simulations with have mass-loss rates in the range
(1.6--4.4)$\times 10^{-10}$\msunyr (see table 1). We have fitted a
``beta'' velocity law to the first 50 points (upto 1.3\rstar) of the
initial conditions and find that they are all well fit by 
$v~=2100(1~-~\rstar\!/r)^{0.8}$\kms, although for the models with
higher mass loss rate, $\beta$ falls to 0.73.
With this velocity structure, we calculate the decoupling velocities
and radii as described in \S2. These are presented in table 1.
Note, we have assumed that all models have solar metallicity for the
calculation of the decoupling radii, which may be inconsistent with
the generation of the initial conditions if the low radiative
acceleration is interpreted as due to a change in metallicity.

\begin{table*}
\caption{Model parameters for the simulations. $r_d$ and $v_d$ are the
decoupling radius and velocity respectively. The stalling radius $r_s$
is calculated from eq.2. The shell period and the
median maximum emission measure (over several cycles) are listed in columns
6 and 7.} 
\begin{tabular}{ccccccc}
Model & log$_{10}\dot{M}$ (\msunyr) &
$r_d$(\rstar) & $v_d$ (\kms)& $r_s$(\rstar)& log$_{10}$EM &
period (hr) \\ \hline
A & -9.35 & 1.27 & 590 & 3.50 & 56.0 & 27.8 \\
B & -9.46 & 1.26 & 568 & 3.09 & 55.7 & 20.8 \\
C & -9.59 & 1.23 & 522 & 2.39 & 55.1 & 14.6 \\
D & -9.68 & 1.20 & 447 & 1.82 & 54.6 &  8.9 \\
E & -9.78 & 1.17 & 378 & 1.53 & 54.1 &  6.6 \\
\end{tabular}
\end{table*}

In order to simulate the decoupled flow, we use these fully coupled
wind structures as initial conditions and simply turn off the
radiation force at the decoupling radius. The flow is calculated in
each simulation for at least ten shell episodes to ensure that the
initial conditions have no effect on the subsequent flow.

\begin{figure}
\begin{picture}(200,380)
\put(0,0){\includegraphics{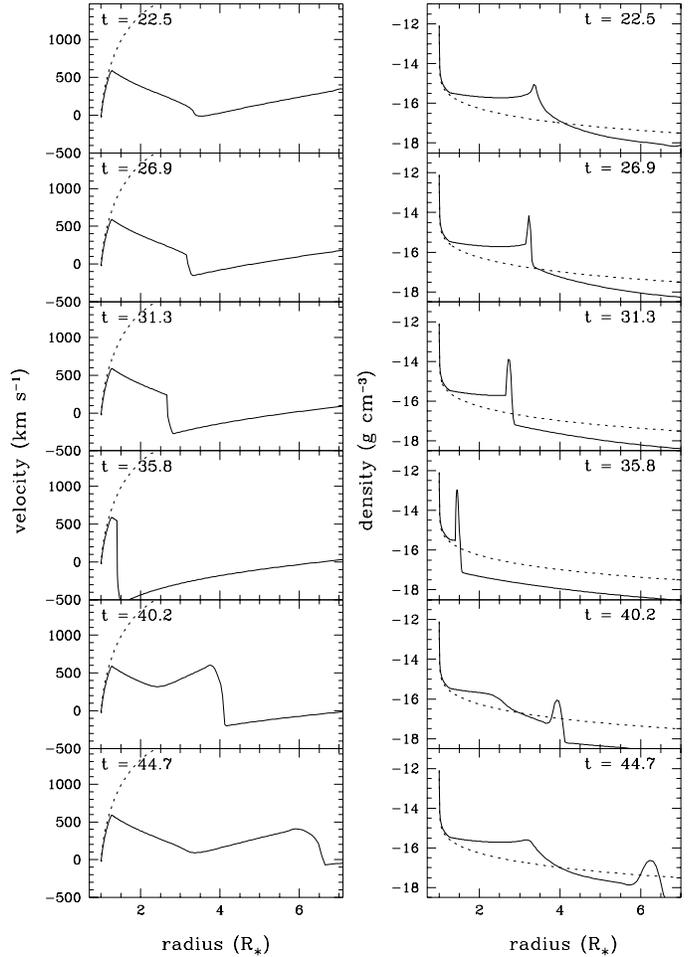}}
\end{picture}
\caption{Numerical hydrodynamic modelling of a the formation and
evolution of a spherical shell (Model E). The dotted line is the steady,
fully-coupled starting condition. The plots are labelled in time (in
hours). The radiation force was set to zero at radii larger than 1.27\rstar.}
\end{figure}

Part of the simulation's results for model A are shown in fig.2 -- the
panels of 
velocity (left) and density (right) are labelled with the time
in hours (the starting point for the time has been shifted to coincide
with fig.3).  
All the models have qualitatively the same behaviour with only the
physical scales changing from model to model.
The first couple of cycles have
not been shown to remove any start-up transients.
The dotted lines on the panels are the starting condition described
above as a reference.
The first panel shows the dense shell forming at a radius of
3.5\rstar$\!$.
The shell's mass increases (to $\sim 1.8\times 10^{-13}\msun$) until
its weight is large
enough to overcome the wind ram-pressure, and so it
contracts toward the star (at around 5hrs later -- the second panel of fig.2).
The shell gains mass and falls faster with time (panels 3 and 4),
finally collapsing onto the star just before 37hrs.
Once the shell makes contact with the star, the wind is again free to
accelerate to decoupling. This process sweeps up some of the material
left behind by the previous cycle's contracting shell and a new
shell is generated moving outwards (the lower two panels of fig.2).
The whole cycle takes around 28hrs. Although the timescale for the
shell growth is very similar to our previous estimate, the
timescale for the whole cycle is $\sim 1$day for this star.
The periods for the shell episodes for the different models are shown
in column 7 of table 1.

We have calculated the emission measure EM$=\int n_e^2 dV$ (where
$n_e$ is the electron density, and $V$ is the volume of the shell) for
the simulations. The median maximum value for the emission measure
from all the simulated cycles is shown in column 6.
Fig.3 shows how the radius, velocity, and emission measure changes
with time through a shell episode. We have identified the
shell at the point where there is a local maximum in density (when the
shell is close to the star it becomes difficult to identify
unambiguously and so these points have been left out of fig.3).
The wind initially sweeps up the ambient gas creating a low
density-contrast shell. 
In this phase the shell is decelerating from the decoupling radius,
and moving out in radius. 
Finally the shell comes to rest ($v_s = 0$ at $r
\approx 3.5$\rstar), and it starts to gain mass. The emission measure starts
to increase significantly during this time. When it cannot be
supported any longer the shell starts to fall back
toward the star, finally coming into contact with the star at a
velocity of around -500\kms. 

All of the simulations produce very similar plots except the scaling
in time and position are different. We find that the stalling radii
are very similar to those predicted using eq.2 (table 1, column 3).
\begin{figure}
\begin{picture}(200,380)
\put(0,0){\includegraphics{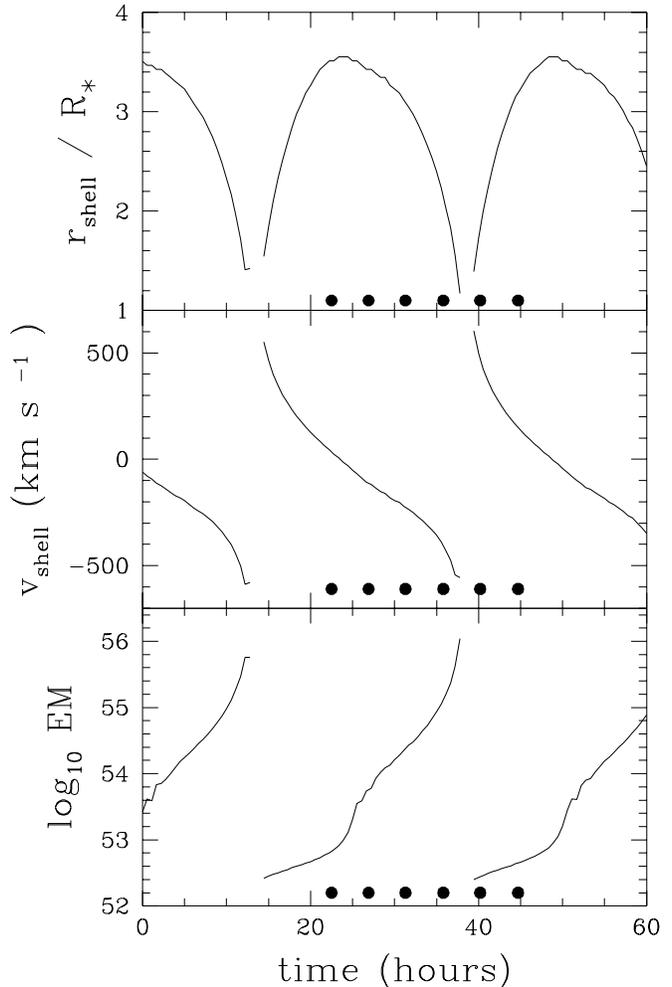}}
\end{picture}
\caption{Time resolved attributes of the shell for Model A. Position
(top) and velocity (middle) of the shell in time
through a shell episode. The lower panel shows the total emission
measure of the wind structure. The solid dots indicate the times of
the corresponding panels in fig.2.}
\end{figure}

The period for initial shell growth has already been shown to be
dependent on the mass-loss rate via the stalling
radius (eq.5) which is, in turn, set by the decoupling radius. 
As the mass-loss rate increases, the decoupling velocity/radius
increases (see table 1), and so the stagnation radius (at which the
shell forms) also increases. 
We expect that the scaling relationships in the shell-growth timescale
(eq.5) is similar for the timescale of the shell episode as a whole,
although the time for shell growth is only a fraction of the shell
cycle. Hence we expect that the period should be proportional to the
stalling radius squared.
In fig.4 the logarithmic period-stalling radius relationship is
displayed. 
\begin{figure}
\begin{picture}(200,250)
\put(0,0){\includegraphics{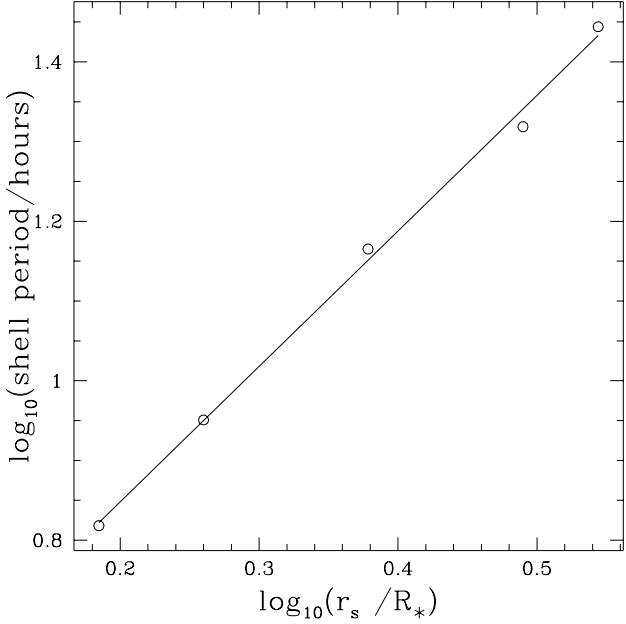}}
\end{picture}
\caption{The log-log relationship between the period of shell episodes
$P$ and the stalling radius
of the wind $r_s$ from eq.2. The straight line is the least-squares
fit to the points: $P = 3.22 (r_s/\rstar)^{1.7}$hours.}
\end{figure}
We have performed a least-squares fit to the above data and find that
the period $P$ is related to the stalling radius via $P = 3.22
r_s^{1.7}$ which is close to the expected scaling. 
We also find that the maximum emission measure follows the same
scaling with radius.

%-----------------------------------------------------------------------%

\section{Observable signatures}

The total mass in the shell will typically be the wind mass-loss rate
multiplied by the total lifetime of the shell. 
Timescales of $\sim$day have been found 
from the previous section, yielding 
maximum shell masses
of $M_{\rm shell} \sim 10^{-13}$--$10^{-12}\msun$. 
Clearly such a low mass will not have a large influence on the gross
observational properties of the star. 
The most obvious effect of decoupling is that the maximum wind velocity
will be observed to be lower than the terminal velocities from
theoretical studies -- a phenomenon which has been known
observationally for some time (Grady \etal\ 1987).

Would the shell give rise to any variation in the lines of the star?
Clearly there would be no change in the UV line profiles until the
shell radius becomes less than the decoupling radius.
Here the maximum edge velocity will
decrease as the shell falls toward the star. 
However, the occurs only over a small fraction of the time of the
whole cycle.

It may be possible that the shell may provide a varying component to
the hydrogen lines.
Let us assume that the shell is always optically thin and calculate
the excess emission at H$_\alpha$. 
This is simply $j = h\nu
\alpha^{\rm eff}_{H_\alpha} n_e n_p / 4\pi$ erg cm$^{-3}$ s$^{-1}$
st$^{-1}$ 
where $n_e$ and $n_p$ are the electron and proton number densities, and
$\alpha^{\rm eff}_{H\alpha}$ is the H$\alpha$
recombination coefficient of $5.96\times 10^{-14}$cm$^3$ s$^{-1}$
assuming Case B recombination at 20,000K (Osterbrock, 1989). 
Assuming that the emission is optically thin then the shell produces
a flat-topped emission profile with a range
in wavelength given by the velocity of the shell. When the shell is
close to the star, then the star occults the far side of the shell and
so curtails the maximum blueshifted emission. 
If the angle between the pole and a point on the shell is $\theta$,
then emission from that point is Doppler shifted by
\begin{equation}
\delta \lambda = \lambda \frac{v_s}{c} \sin\theta
\end{equation}
where $v_s$ is the velocity of the shell, and $c$ is the speed of light.
Combining this with the expression for $j$, and integrating the volume
emitting at the same wavelength, we derive the
emission as
\begin{equation}
j = \frac{h \alpha^{\rm eff}_{H_\alpha} n_e n_p}{2}
r_s^2 dr \left(\frac{c}{v_s}\right) \ \ {\rm erg\ s}^{-1}{\rm
st}^{-1}{\rm hz}^{-1}
\end{equation}

We assume that that star is emitting as a black-body 
and calculate the fraction of the black body flux the
shell emits (of course in a {\em real} star there is also photospheric
absorption).
After some simple manipulation we find that this
ratio is
\begin{equation}
{\cal R} = \frac{\left(e^{h\nu/kT} - 1 \right)}{4} 
\frac{c^3}{\nu^3} \frac{\alpha^{\rm eff}_{H_\alpha} EM}{4\pi\rstar\!^2
v_s}
\end{equation}
where the symbols have their usual meanings. 
This extra component extends over a
wavelength range
\begin{equation}
\lambda_0 \left(1 - \frac{v_s}{c}\sqrt{1 -
\left(\frac{\rstar}{r_s}\right)^2} \right) < \lambda <
\lambda_0 \left(1 + \frac{v_s}{c} \right)
\end{equation}
where $\lambda_0 = 6563$\AA\ is the central wavelength of the line.
\begin{figure}
\begin{picture}(200,380)
\put(0,0){\includegraphics{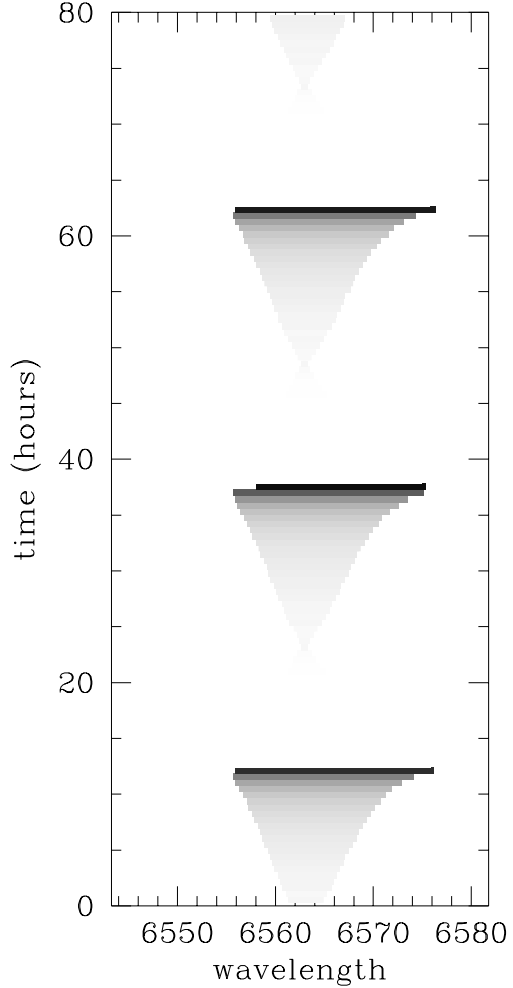}}
\end{picture}
\caption{Grey-scale representation of the time series of the excess
line emission at H$_\alpha$ due to the shell for Model A (see text).
The grey scale in linear, and the darkest tones correspond to 1\% of
the continuum emission.}
\end{figure}

We have calculated the extra component of H$_\alpha$ for Model A and
present it as a time series in fig.5. The grey scale is linear with
a range of 0-1\% of the continuum with the 
the darkest tone corresponding to 1\%.
We find that the maximum value of ${\cal R}$ for Model A is $\sim 1\%$,
i.e. the flat-topped shell component has a contrast of $\sim 1$\% of
the continuum flux.
Whilst this is clearly an observable amount, it is only present
at significant values for a small of the cycle.

%-----------------------------------------------------------------------%
\section{Discussion \& Conclusions}

We have demonstrated that low density radiation-driven flows may
decouple before the wind becomes unbound from the star, and therefore
will reaccrete. 
Abbott \& Friend (1989) discussed a model of a line-driven
wind with a line force cut-off to mimic the effects of shocks in the
wind. However, their models had force cut-offs such that the wind
beyond decoupling was unbound and therefore would escape from the star. 

This is the first suggestion that these flows may still be bound to
the star post-decoupling. 
We have examined the attributes of the periodic shell
structures formed in the flow and have calculated the line emission.

From our analysis we conclude that these shells may be observable in a
line monitoring campaign, with the largest emission coming from stars
with winds which decouple close to the escape velocity. These stars
will also have the longest period for the shells making it more
probable that the shell can be observed.

It is worthy of note that this process of reaccretion will change the
metallicity in the outer layers of the star over a long time
period. As it is only the metals which become unbound, then their
abundance decreases with time in the outer layers when the shells
reaccrete. 
In a future study we will address the case of rotating winds which may
be applicable to classical Be stars as these stars do seem to have
small mass-loss rates.

%-----------------------------------------------------------------------%

\section*{Acknowledgements}
JMP is supported by a PPARC postdoctoral research assistantship.
The referee, Dr. S.P. Owocki is thanked for constructive suggestions
which improved this paper.

%-----------------------------------------------------------------------%

%-----------------------------------------------------------------------%
\end{document}